# Divergent Evolution of Progesterone and Mineralocorticoid Receptors in Terrestrial Vertebrates and Fish Influences Endocrine Disruption

Michael E. Baker
Division of Nephrology-Hypertension
Department of Medicine, 0735
University of California, San Diego
9500 Gilman Drive
La Jolla, CA 92093-0735

Center for Academic Research and Training in Anthropogeny (CARTA)
University of California, San Diego
La Jolla, CA 92093

Correspondence to: mbaker@ucsd.edu

**Abstract:**

There is much concern about disruption of endocrine physiology regulated by steroid hormones in humans, other terrestrial vertebrates and fish by industrial chemicals, such as bisphenol A, and pesticides, such as DDT. These endocrine-disrupting chemicals influence steroid-mediated physiology in humans and other vertebrates by competing with steroids for receptor binding sites, disrupting diverse responses involved in reproduction, development and differentiation. Here I discuss that due to evolution of the progesterone receptor (PR) and mineralocorticoid receptor (MR) after ray-finned fish and terrestrial vertebrates diverged from a common ancestor, each receptor evolved to respond to different steroids in ray-finned fish and terrestrial vertebrates. In elephant shark, a cartilaginous fish, ancestral to ray-finned fish and terrestrial vertebrates, both progesterone and 17,20β-dihydroxy-progesterone activate the PR. During the evolution of ray-finned fish and terrestrial vertebrates, the PR in terrestrial vertebrates continued responding to progesterone and evolved to weakly respond to 17,20β-dihydroxy-progesterone. In contrast, the physiological progestin for the PR in zebrafish and other ray-finned fish is 17,20β-dihydroxy-progesterone, and ray-finned fish PR responds weakly to progesterone. The MR in fish and terrestrial vertebrates also diverged to have different responses to progesterone. Progesterone is a potent agonist for elephant shark MR, zebrafish MR and other fish MRs, in contrast to progesterone's opposite activity as an antagonist for aldosterone, the physiological mineralocorticoid for human MR. These different physiological ligands for fish and terrestrial vertebrate PR and MR need to be considered in applying data for their disruption by chemicals in fish and terrestrial vertebrates to each other.

**Introduction**

Adrenal and sex steroid receptors belong to the nuclear receptor family, a diverse group of transcription factors that arose in multicellular animals [1–3]. These ligand-activated transcription factors include the estrogen receptor (ER), androgen receptor (AR), progesterone receptor (PR), mineralocorticoid receptor (MR) and glucocorticoid receptor (GR) [1,4–6]. Binding of steroids (Figure 1) to their cognate receptors initiates a complex process involving steroid-dependent conformational changes in the receptor and the binding of the hormone-receptor complex to co-regulator proteins and specific hormone-response elements on DNA in target cells, which leads to transcription of genes important in diverse physiological pathways involved in differentiation, development, reproduction, immune responses, homeostasis and stress responses in vertebrates [7–12].

This special issue focusses on endocrine disruption of steroid receptors, which is a major concern for the health of humans and fish and other wild-life [13–17]. Endocrine disruption is a consequence of our modern industrial society and the explosive increase in the volume and variety of synthetic chemicals that are released into the environment, which then disrupt steroid hormone physiology by interfering with the activation of steroid receptor mediated transcription. Studies of the effects of synthetic chemicals on the physiology of zebrafish [18–25] have been useful in deciphering the biological effects and the mechanism of action of many chemicals and drugs for use in evaluating their effects on human health. Compared to mice, zebrafish are relatively inexpensive to maintain, and they can be bred in large numbers. An important advantage of zebrafish embryos is that they are optically transparent, which allows microscopic observations of organs as they develop.

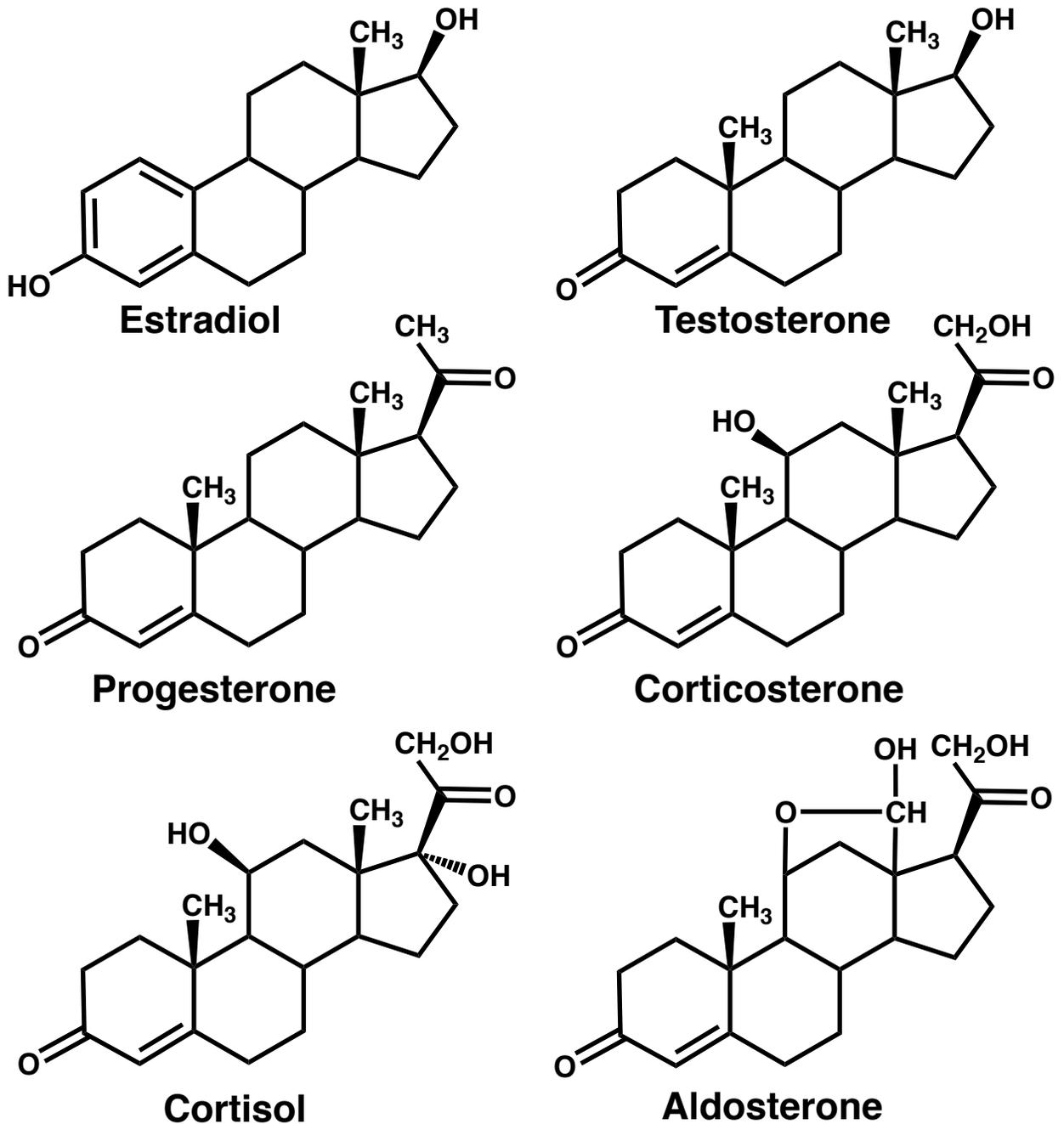

**Figure 1. Structures of vertebrate adrenal and sex steroids.**
In terrestrial vertebrates, cortisol, corticosterone, aldosterone, progesterone, estradiol and testosterone are the major circulating adrenal and sex steroids. Estradiol is important in female and male physiology [26]. Aldosterone is the main physiological mineralocorticoid in terrestrial vertebrates [27,28]. Cortisol and corticosterone are physiological glucocorticoid in vertebrates [5,8]. Cortisol also activates the MR, although at about a 10-fold higher steroid concentration than aldosterone [29,30]. Testosterone and progesterone are reproductive hormones in males and females, respectively. Progesterone is an antagonist for the MR in humans [31], alligators and *Xenopus* [29] and an agonist for fish MRs [29,32,33].

Here I use an evolutionary perspective to discuss a largely ignored difference between human and fish in the responses of their PR and MR to different physiological steroids, and this, I propose, indicates the need for caution in extrapolating data about chemical disruption of the PR and MR in fish to human physiology and vice versa. Other hormone receptors, e.g. androgen receptor, in terrestrial vertebrates and fish also have different physiological ligands, and, thus, provide an inaccurate assumption that some chemicals are endocrine disruptors for both fish and terrestrial vertebrates.

**Steroid receptors evolved about 550 million years ago.**

Multicellular animals evolved about 1 billion years ago [1,34–38]. Orthologs of the ER and a 3-keto-receptor receptor (SR) appear about 550 million years ago in an ancestor of modern amphioxus, a cephalochordate that evolved at the base of the vertebrate line [1,35,39–42]. The 3-keto-steroid-receptor is the ancestor of the PR, MR, GR and AR [35,39,40]. An ortholog of the PR evolved in ancestor of modern sea lampreys and hagfish, which are jawless fish (Cyclostomes) [43–45]. A corticoid receptor (CR), which is the ancestor of the MR and GR, also evolved in an ancestral jawless fish [1,5,34,35,40]. An AR and a distinct GR and MR evolved in cartilaginous fishes, ancestors of sharks, skates and rays, which are jawed vertebrates that diverged from bony vertebrates about 450 million years [5,46]. Land animals and ray-finned fish diverged about 425 million years ago [47,48]. Here I discuss the implications for endocrine disruption of the PR and MR in ray-finned fish and terrestrial vertebrates, which evolved to respond to different physiological steroids after their split about 425 million years ago [47,48].

**Different Physiological Steroids are Transcriptional Activators of the Progesterone Receptor in Ray-Finned Fish and Terrestrial Vertebrates**.

The physiological steroid for transcriptional activation of human PR is progesterone [49–51] (Figure 1), which has a half maximal response (EC50) of 0.12 nM for human PR (Table 1). Human PR has little activity for progesterone analogs containing a 17α-hydroxyl group (Figure 2, Table 1). Thus, addition of a 17α-hydroxyl group to form 17αOH-Progesterone (17αOH-Prog) (Figure 2) increases the EC50 to 132 nM and addition of a second hydroxyl at C20 to form 17α,20β-OH-Progesterone (17α,20β-P, 17α,20β-OH-P) (Figure 2) increases the EC50 to 344 nM for human PR.

**Table 1. EC50 values for steroid activation of elephant shark PR, human PR and zebrafish PR.**
**A. Progesterone, 20β-OH-Progesterone, 17α-OH-Progesterone.**

|  | **Progesterone** | **20β-OH-Progesterone** | **17α-OH-Progesterone** |
|---|---|---|---|
|  | **EC50 (M)** | **EC50 (M)** | **EC50 (M)** |
| **Elephant shark PR** | **0.18 nM | **0.48 nM | **0.36 nM |
| **Human PR** | **0.13 nM | **4.6 nM | **113.0 nM |
| **Zebrafish PR** | *296.2 nM | Not Determined | *193.4 nM |

*Chen et al. Biology of Reproduction 82, 171–181 (2010)
**Lin et al. bioRxiv, 2021 doi: https://doi.org/10.1101/2021.01.20.427507, (n.d.).

**B. Fish progestins: 17α,20β-dihydroxy-4-pregnene-3-one, 17α,20β,21-trihydroxy-4-pregnen-3-one.**

|  | **17α,20β-dihydroxy-4-pregnene-3-one**<br>**17α,20β-DHP, 17α,20β-P** | **17α,20β,21-trihydroxy-4-pregnen-3-one**<br>**20β-S, 17,20β,21-P** |
|---|---|---|
|  | **EC50 (M)** | **EC50 (M)** |
| **Elephant shark PR** | **2.6 nM | **34.4 nM |
| **Human PR** | **408 nM | **Not active |
| **Zebrafish PR** | *8 nM | *40.9 nM |

*Chen et al. Biology of Reproduction 82, 171–181 (2010)
**Lin et al. bioRxiv, 2021 doi: https://doi.org/10.1101/2021.01.20.427507, (n.d.).

**C. Corticosteroids: Cortisol, 11-deoxycortisol, Corticosterone, DOC.**

|  | **Cortisol** | **11-deoxycortisol** | **Corticosterone** | **DOC** |
|---|---|---|---|---|
|  | **EC50 (M)** | **EC50 (M)** | **EC50 (M)** | **EC50 (M)** |
| **Elephant shark PR** | **52.4 nM | **0.47 nM | **10.5 nM | **0.19 nM |
| **Human PR** | **826 nM | **39 nM | **8.2 nM | **1.4 nM |
| **Zebrafish PR** | *Inactive at 1uM | Not Determined | Not Determined | Not Determined |

*Chen et al. Biology of Reproduction 82, 171–181 (2010)
**Lin et al. bioRxiv, 2021 doi: https://doi.org/10.1101/2021.01.20.427507, (n.d.).

**Abbreviations:**
**Fish progestins: 20β-OH-Prog = 20β-hydroxy-progesterone.**
**17α,20β-DP, 17α,20β-P = 17α,20β-dihydroxy-4-pregnene-3-one,**
**20β-S, 17,20β,21-P = 17α,20β,21-trihydroxy-4-pregnen-3-one.**
**Corticosteroids: DOC=11-deoxycorticosterone.**

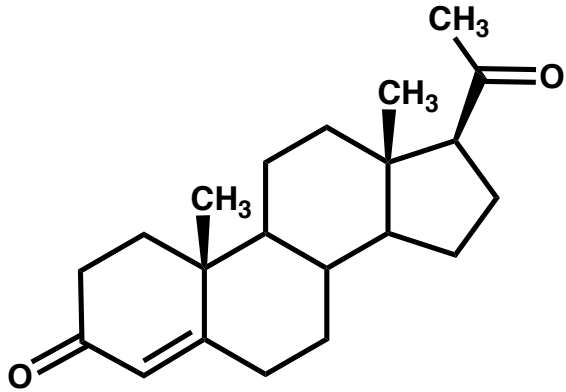
**Progesterone**

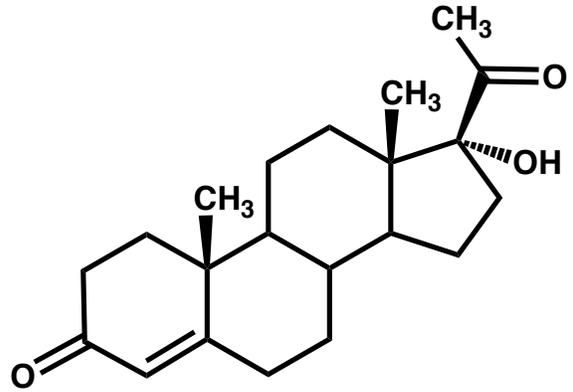
**17-OH-Progesterone**

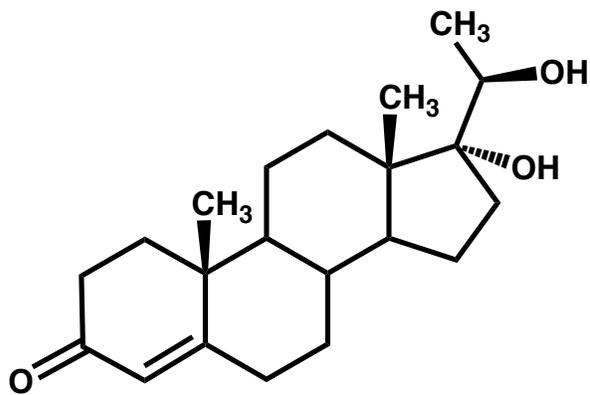
**17α,20β-dihydroxy-4-pregnen-3-one**
**17α,20β-OH-Progesterone**

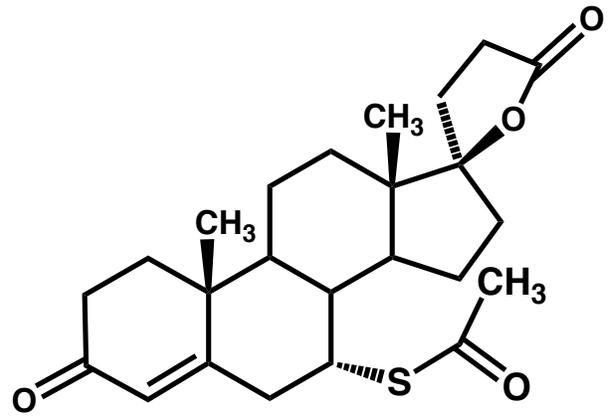
**Spironolactone**

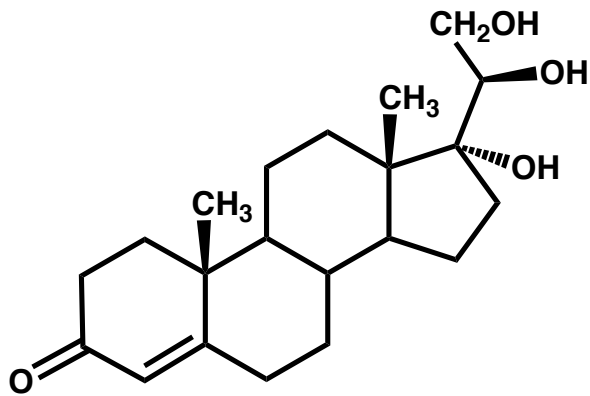
**17α,20β,21-trihydroxy-4-pregnen-3-one**
**17α,20β,21-OH-Progesterone**

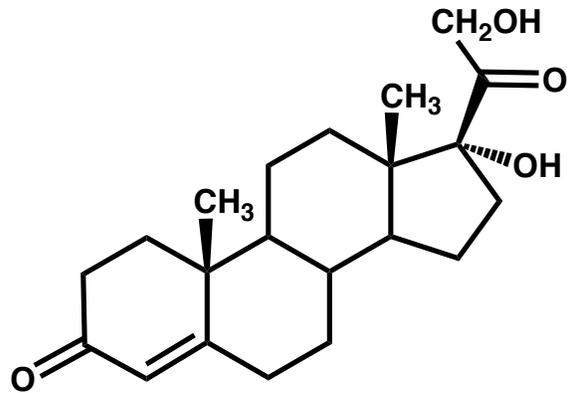
**11-Deoxycorticosterone**
**21-OH-Progesterone**

**Figure 2. Structures of steroids that activate the progesterone receptor and mineralocorticoid receptors in elephant shark, ray-finned fish and humans.**
The physiological progestins for the PR in zebrafish and other ray-finned fish are 17,20β-dihydroxy-progesterone and 17,20β,21-trihydroxy-pregn-4-en-3-one, which also activate elephant shark PR (Table 2) [52]. Ray-finned fish PR respond weakly to progesterone [52–54], while human PR and elephant shark PR are activated by progesterone [55]. Spironolactone, which has structural similarities to progesterone activates ray-finned fish PR and elephant shark PR [55], while spironolactone inhibits transcriptional activation of human PR [31]. Progesterone and spironolactone activate elephant shark MR and ray-finned fish MR and inhibit transcriptional activation of human MR.

Similar to human PR, elephant shark PR has a low EC50 (0.18 nM) for progesterone (Table 1). However, compared to human PR, the response to hydroxylated progesterone analogs is very different in elephant shark PR, which has low EC50s for 17α-OH-Prog (EC50=0.36 nM) and 17α,20β-P (EC50=2.6 nM) (Figure 2) (Table 1) [55], two steroids that have little activity for human PR (Table 1). Like elephant shark PR, fish PR has a strong response to 17,20β-P, which is important in fish reproductive physiology, including acting as a maturation inducing hormone of teleost fish [52,56–60]. Indeed, 17,20β-P is the major physiological progestin in fish, although in some fish, 17,20β,21-trihydroxy-pregn-4-en-3-one (17,20β,21-P, 20β-S,) is a physiological progestin [52–54,61]. Importantly, progesterone is a weak activator of the PR in zebrafish PR [59] (Table 1) and other fish [52–54].

Thus, after the split of ray-finned fish and terrestrial vertebrates from an ancestral cartilaginous fish, their PRs evolved to respond to different steroids.

**Progesterone is a Transcriptional Activator of Fish MR and a Transcriptional Inhibitor of the Human MR.**

Although aldosterone is the physiological mineralocorticoid in humans and other terrestrial vertebrates [27,28,62–67], cortisol and corticosterone (Figures 1, 2) also are transcriptional activators of human MR. Among these steroids, aldosterone has the lowest EC50 (Table 2). Moreover, despite the high affinity of progesterone for human MR [31,68–70], progesterone is an antagonist for human MR [31,69,71]. Interestingly, spironolactone, an anti-mineralocorticoid that is used clinically to inhibit aldosterone activation of human MR, has structural similarities to progesterone (Figure 2) [31,69,72,73].

**Table 2. EC50 values for activation by progestins, spironolactone and corticosteroids of elephant shark MR, human MR and zebrafish MR.**

**A. Progesterone, 19-norProgesterone, 17-OH-Progesterone and Spironolactone activation of elephant shark MR, human MR and zebrafish MR.**

|  | Progesterone | 19-norProg | 17-OH-Prog | Spironolactone |
|---|---|---|---|---|
|  | EC50 (M) | EC50 (M) | EC50 (M) | EC50 (M) |
| Elephant shark MR | **0.45 nM <br> *0.27 nM | **0.11 nM <br> *0.043 nM | *1.4 nM | *0.55 nM |
| Human MR | Not active | Not active | Not active | Not active |
| Zebrafish MR | *2.4 nM | *0.94 nM | *18 nM | *3.8 nM |

**Katsu et al. J. Steroid Biochemistry & Molecular Biology, 210, 2021.
*Katsu et al. Sci Signal.2019, 12(584):eaar2668. doi: 10.1126/scisignal.aar2668.

**B. Corticosteroid activation of elephant shark MR, human MR and zebrafish MR.**

|  | Aldosterone | Cortisol | Corticosterone |
|---|---|---|---|
|  | EC50 (M) | EC50 (M) | EC50 (M) |
| Elephant shark MR | **0.14 nM <br> *0.11 nM | **1.6 nM <br> *0.46 nM | **0.61 nM <br> *0.17 nM |
| Human MR | #0.27 nM | #5.5 nM | #1.2 nM |
| Zebrafish MR | #0.082 nM | #0.44 nM | #0.3 nM |

|  | DOC | 11-deoxycortisol |
|---|---|---|
|  | EC50 (M) | EC50 (M) |
| Elephant shark MR | **0.1 nM <br> *0.06 nM | **0.2 nM <br> *0.11 nM |
| Human MR | #0.42 nM | #3.6 nM |
| Zebrafish MR | #0.06 nM | #0.4 nM |

**Katsu et al. J. Steroid Biochemistry & Molecular Biology, 210, 2021.
*Katsu et al. Sci Signal.2019, 12(584):eaar2668. doi: 10.1126/scisignal.aar2668.
#Katsu et al. Sci Signal. 2018 Jul 3;11(537):eaao1520. doi: 10.1126/scisignal.aao1520.

However, instead of inhibiting fish MR, progesterone is a strong activator of the MR in trout [32], sturgeon [33], gar [33] and zebrafish [29,69,74], while spironolactone activates the MR in trout [32] and zebrafish [29,69,74,75], as well as in elephant shark MR [74,76]. Activation by progesterone and spironolactone of the MR in ray-finned fish conserves the MR response in cartilaginous fish [74,76], while, in a mammalian ancestor, the MR [69,71,77] diverged to be inhibited by progesterone and spironolactone. Interestingly, activation of the MR by progesterone and spironolactone is conserved in lungfish [69,71,77,78], in which synthesis of aldosterone evolved [71,79–81]. It is in amphibians that the mutation in the MR that conferred antagonist activity for progesterone and spironolactone evolved [69,71,77].

These differences between the response in humans and fish of the PR and MR to steroids indicates that some responses of the PR and MR to environmental chemicals will differ between human and fish. This difference also may occur when using zebrafish for testing new drugs for treating cancer and other diseases in humans [18,21–24].

Regarding the physiological role of zebrafish MR, unexpectedly, the MR does not appear to regulate sodium uptake, the classical "mineralocorticoid function", in zebrafish, although the MR is expressed in kidney and gill. Instead, cortisol regulates sodium uptake in zebrafish through transcriptional activation of the GR [82–84], indicating that the MR has other functions in fish kidney and gill, as well as in non-traditional organs, such a heart and brain [63,64,73,84–86]. Thus, differences in functions of the MR and GR in humans and zebrafish also need to be considered when testing chemicals for endocrine disruption and when developing pharmaceuticals for human health.

Lastly, divergent evolution of other steroid receptors in fish and terrestrial vertebrates may also be relevant for different responses to xenobiotics by steroid receptors in humans and fish. For example, in humans, testosterone and dihydrotestosterone are the physiological androgens. In fish, 11-ketotestosterone is an important androgen [87–89]. Analysis of transcriptional activation by androgens and other steroids in fish, humans and elephant sharks may uncover additional evolutionary insights that have practical application to our health.

**Funding, Contributions and Competing Interests.**

**Funding:** M.E.B. was supported by UC San Diego Foundation Research fund #3096.

**Author contributions:** M.E.B. wrote the paper.

**Competing Interests:** M.E.B. has no competing interests.